\documentclass[journal]{IEEEtran}

\usepackage{graphicx}
\usepackage{lineno}
\usepackage{stfloats}
\usepackage{xcolor}
\usepackage{ragged2e}

\ifCLASSINFOpdf

\else

\fi

\usepackage{graphicx}
\usepackage{lineno}
\usepackage{stfloats}
\usepackage{multicol}
\usepackage{array}
\usepackage{verbatim}

\begin{document}

\title{Compact and fully functional high-frequency sine wave gating InGaAs/InP single-photon detector module}

\author{Qi~Xu, 
	    Chao~Yu,
	    Dajian~Cui, 
	    Xuan-Yi~Zhang,
	    Wei~Chen, 
	    Yu-Qiang~Fang,
	    Lianjun~Jiang,
	    Qixia~Tong,
	    Jianglin~Zhao,
	    and Jun~Zhang
       
\thanks{Qi Xu, Chao Yu and Xuan-Yi Zhang are with Hefei National Research Center for Physical Sciences at the Microscale and School of Physical Sciences, University of Science and Technology of China, Hefei 230026, China, and are also with CAS Center for Excellence in Quantum Information and Quantum Physics, University of Science and Technology of China, Hefei 230026, China (e-mail: yuch@ustc.edu.cn). }
\thanks{Dajian Cui, Wei Chen, Qixia Tong and Jianglin Zhao are with Chongqing Optoelectronics Research Institute, Chongqing 400060, China, and are also with Chongqing Key Laboratory of Core Optoelectronic Devices for Quantum Communication, Chongqing 400060, China.}
\thanks{Yu-Qiang Fang and Lianjun Jiang are with QuantumCTek Co., Ltd., Hefei 230088, China.}
\thanks{Jun Zhang is with Hefei National Research Center for Physical Sciences at the Microscale and School of Physical Sciences, University of Science and Technology of China, Hefei 230026, China, and with CAS Center for Excellence in Quantum Information and Quantum Physics, University of Science and Technology of China, Hefei 230026, China, and also with Hefei National Laboratory, University of Science and Technology of China, Hefei 230088, China (e-mail: zhangjun@ustc.edu.cn). }}
\maketitle

\begin{abstract}
High-frequency sine wave gating (SWG) InGaAs/InP single-photon detectors (SPDs) are widely used for synchronous near-infrared single-photon detection. For practical use, the size of SPD is one of the most concerning features for system integration. Here we present, to the best of our knowledge, the most compact and fully functional high-frequency SWG InGaAs/InP SPD. We develop a sine wave gating integrated circuit (SWGIC) using system-in-package technology that supports functions including large amplitude sine wave gate generation, coincidence gate generation, phase regulation, amplitude monitoring, and amplitude modulation. Moreover, we design and fabricate a high-performance multi-mode fiber coupled InGaAs/InP single-photon avalanche diode (SPAD) with a compact butterfly package. Furthermore, we implement a monolithically integrated readout circuit (MIRC) to extract the weak avalanche signal from large capacitance response of SWG. Finally, the SWGIC, SPAD, MIRC, and the affiliated circuits are integrated into a single module with a size of 6 cm $\times$ 5.7 cm $\times$ 1.7 cm. After characterization, the SPD module exhibits a photon detection efficiency of 40\%, a dark count rate of 9 kcps, and an afterpulse probability of 4.6\% at an operation temperature of 238 K and a hold-off time of 160 ns. Our work provides a practical solution for applications necessitating highly integrated near-infrared single-photon detection.
\end{abstract}


\section{Introduction} 
\IEEEPARstart
InGaAs/InP single-photon detectors (SPDs) are practical candidates for near-infrared single-photon detection due to their small size, low cost, and ease-of-operation\cite{ZJ15review,YC24Rev}. Therefore, they have been widely used in applications such as quantum key distribution (QKD)\cite{Rev21QP}, light detection and ranging (Lidar)\cite{Rev23SPDforlidar}, and fluorescence lifetime imaging\cite{Rev20FLIM}. InGaAs/InP SPDs can be operated in either the free-running or gating mode. Free-running mode detectors are versatile owing to their ability to detect photons arriving at any moment. Free-running InGaAs/InP SPDs can be implemented using various methods\cite{JXD09NFAD,Geneva14LDCR,LYF21AQ,WL23PQ,ZJ23PDE}. However, they suffer from relatively high afterpulse probability, thus a long hold-off time usually needs to be set, which results in a relatively low maximum count rate.

High-frequency gating technique provides a high-performance solution for synchronous photon detection, which is highly desired in QKD applications\cite{ZJ15review,YC24Rev}. Compared to free-running mode SPDs, high-frequency gating SPDs exhibit a lower dark count rate (DCR) due to their small duty cycle. The narrow gate significantly mitigates the avalanche time, leading to a considerably lower afterpulse probability. Furthermore, the maximum count rate of high-frequency gating SPDs is significantly higher than that of free-running mode SPDs, reaching several hundreds of Mcps\cite{ZHP23MCR,YZL23}. High-frequency gating SPDs can also be used for asynchronous photon detection, called gated-free mode, to obtain a high dynamic range at the cost of photon detection efficiency (PDE) loss\cite{Milan13GateFree}. For example, high-frequency gating SPDs have been employed in numerous Lidar applications\cite{ZHP14imaging,ZHP15CAM,Jap20imaging}.

The primary techniques for realizing high-frequency gating include self-differencing\cite{YZL15}, capacitance-balancing\cite{Milan151G3gate,Park19DA}, harmonic subtraction\cite{Res16}, and sine-wave gating (SWG)\cite{Nihon06SWG,ZJ09ASIC,ZJ20PDE}. Among them, SWG is the most widly utilized owing to its stability and ease of implementation. Generally, SWG InGaAs/InP SPDs comprise SWG circuits, single-photon avalanche diodes (SPADs), readout circuits, and affiliated control circuits. For practical applications, the size of SPD is one of the most concerning features for system integration. In 2012, Liang $et$ $al.$\cite{ZJ121G25} presented a fully integrated 1.25 GHz SWG InGaAs/InP SPD with a size of 25 cm $\times$ 10 cm $\times$ 33 cm. In 2018, by employing a monolithic integrate readout circuit (MIRC) and a butterfly package SPAD with a mini-thermoelectric cooler inside, Jiang $et$ $al.$\cite{ZJ18MSWG} decreased the SPD size to 13 cm $\times$ 8 cm $\times$ 4 cm. In 2023, Yan $et$ $al.$\cite{YZL23Compact} reported a compact 1.25 GHz SWG InGaAs/InP SPD with a size of 8.8 cm $\times$ 6 cm $\times$ 2 cm. However, it requires an external sine wave seed signal with an appropriate frequency, amplitude, and phase. In other words, the proposed SPD cannot be independently operated.

In this paper, we present a plug and play 1.25 GHz SWG InGaAs/InP SPD with a size of 6 cm $\times$ 5.7 cm $\times$ 1.7 cm. The SPD size is minimized through several techniques. First, a system-in-package sine wave gate integrated circuit (SWGIC) is developed to generate a stable 1.25 GHz sine wave gate with adjustable amplitude and phase. Then, a butterfly-packaged InGaAs/InP SPAD coupled with a 62.5 $\mu$m multi-mode fiber is designed and fabricated. Subsequently, a novel MIRC chip is developed to extract weak avalanche signals. Finally, the SWGIC, SPAD, MIRC, and affiliated circuits are integrated into a single module with a size of 6 cm $\times$ 5.7 cm $\times$ 1.7 cm. After characterization, the minimized 1.25 GHz SWG InGaAs/InP SPD module simultaneously achieves a PDE of 40\%, a DCR of 9 kcps, and an afterpulse probability of 4.6\% at an operation temperature of 238 K and a hold-off time of 160 ns.

\section{1.25 GHz Sine Wave Gating Integrated Circuit}

\begin{figure*}[htbp]
\centerline{\includegraphics[width=17 cm]{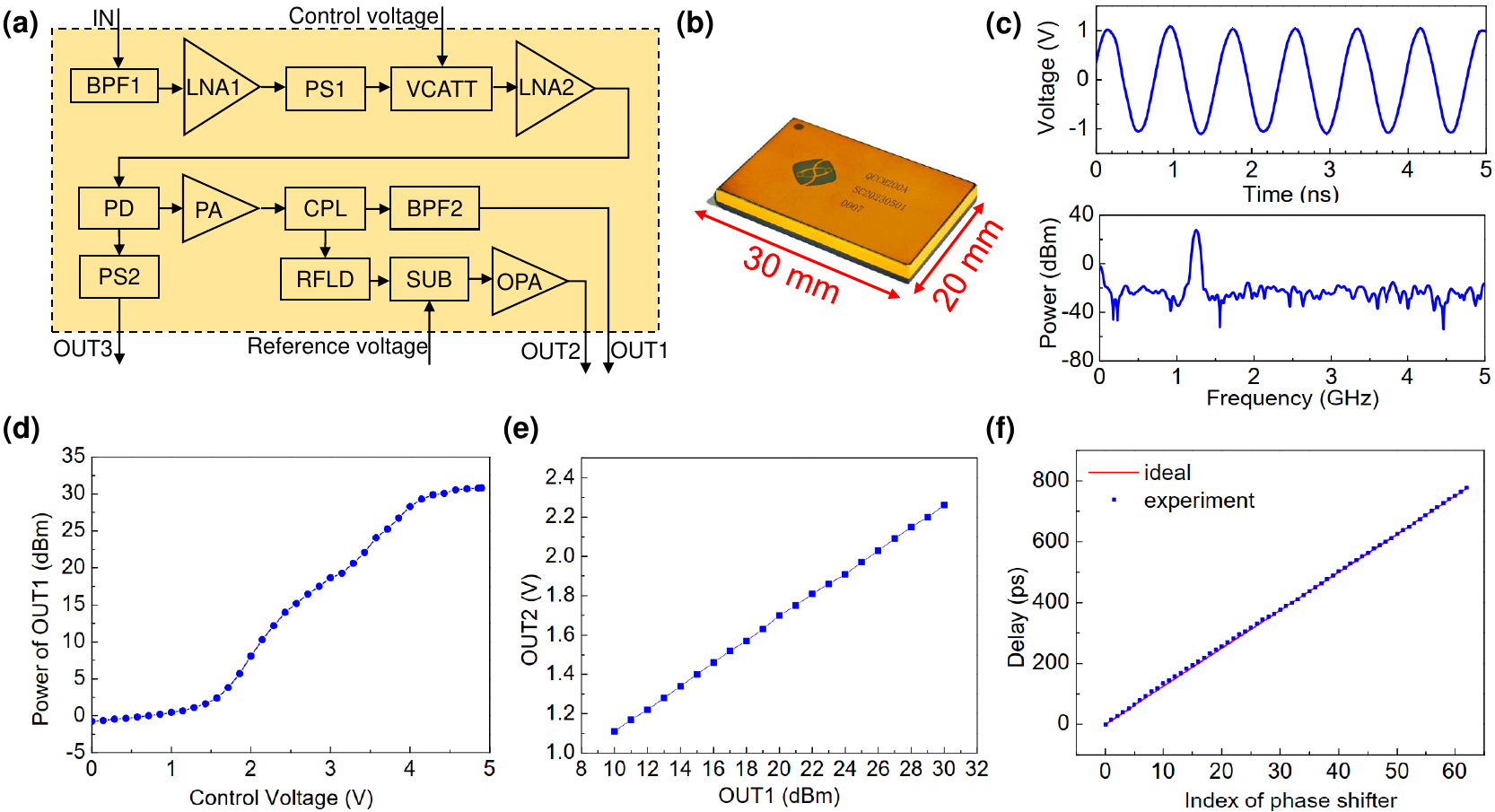}}
\caption{(a) SWGIC diagram. BPF: band-pass filter. LNA: low-noise amplifier. PS: phase shifter. VCATT: voltage-controlled attenuator. PD: power divider. PA: power amplifier. CPL: coupler. RFLD: radio frequency logarithmic detector. SUB: subtractor. (b) Photo of the SWGIC. (c) Typical gate signal output from OUT1 of the SWGIC. (d) Variation of the output power of OUT1 with the control voltage. (e) Output power of OUT1 monitored in real-time by OUT2. (f) Ideal and experimentally measured delays of the gate signal as a function of the PS1 index.}
\label{figure1}
\end{figure*}

The SWG circuit is a complex and crucial part of high-frequency SWG SPDs, with its primary function being high-amplitude sine wave gate generation. To optimize the overall performance of the SPD, the gate amplitude must be adaptable to different operating conditions\cite{ZJ20PDE}. In practical applications such as QKD, precise phase regulation of the sine wave gate is essential for ensuring that photons are detected at the point of maximum detection efficiency. Moreover, the gate amplitude needs to be monitored in real time to maintain long-term stability of the SPD performance. Due to timing jitter, a few parts of the detection signals fall during the gate-off duration, thus a synchronized coincidence signal is usually required to eliminate these noises\cite{ZJ121G25}. In summary, an SWG circuit must support functionalities including gate amplitude modulation, phase regulation, gate amplitude monitoring, and coincidence signal generation.

We implement the aforementioned functions in a SWGIC, the schematic diagram is illustrated in Fig.~\ref{figure1} (a). A 1.25 GHz sine wave signal is employed as the input. In the SWGIC, the input signal first passes through a band-pass filter (BPF1), and is pre-amplified by a low noise amplifiers (LNA1). Then, the phase and amplitude of the pre-amplified signal are regulated using a 6-bit digital controlled phase shifter (PS1) and a voltage controlled attenuator (VCATT). Subsequently, the signal is further amplified by a second low-noise amplifier (LNA2). The total gain of the two LNAs is around 35 dB. Thereafter, the signal is divided into two channels by a power divider (PD). One channel is regulated by a 6-bit phase shifter (PS2) and serves as the coincidence signal output (OUT3). The other channel is amplified by a power amplifier with a gain of 30 dB and divided into two channels by a microstrip coupler (CPL). One channel passes through a band-pass filter (BPF2) and outputs as the gate signal (OUT1). The other channel is detected by a radio frequency logarithmic detector (RFLD). The RFLD output is compared to a reference voltage, and after amplification by an operational amplifier (OPA), the gate amplitude monitoring signal (OUT2) is afforded. All the above mentioned components are integrated into the SWGIC with dimensions of 30 mm $\times$ 20 mm $\times$ 3 mm. Fig.~\ref{figure1} (b) depicts a photograph of the SWGIC.

The typical gate signal output from OUT1 of the SWGIC is shown in Fig.~\ref{figure1} (c), measured using an oscilloscope after applying 20 dB attenuation. In the time domain, the signal exhibits a clean sinusoidal waveform with a peak-to-peak voltage (Vpp) of 2 V, corresponding to a 20 V Vpp at the OUT1 output. In the frequency domain, the signal's power peaks at 1.25 GHz, with a full width at half maximum (FWHM) of 100 MHz. When the control voltage of the VCATT is increased from 0 to 5 V, the output power of OUT1 monotonically rises, with the maximum power reaching $\sim$ 30.8 dBm, corresponding to a 21.9 V Vpp. The measured results are presented in Fig.~\ref{figure1} (d). 

The output power of OUT1 is monitored in real-time by OUT2. Fig.~\ref{figure1} (e) illustrates the relationship between the power at OUT1 and the corresponding voltage at OUT2. When the output power of OUT1 increases from 10 to 30 dBm, the output voltage of OUT2 linearly increases from 1.1 to 2.2 V. Moreover, the overall phase of the gate and coincidence signals is controlled by PS1, while the relative phase between the two signals is regulated by PS2. Since the system's period is 800 ps and both PS1 and PS2 have a maximum index of 64, the minimum delay adjustment step is 12.5 ps. Fig.~\ref{figure1} (f) presents the ideal and experimentally measured delays of the gate signal as a function of PS1 index, the maximum error of delay is 9 ps, while the mean error is 3 ps. The performance of PS2 exhibits a similar accuracy level.

\section{InGaAs/InP SPAD}

\begin{figure*}[htbp]
\centerline{\includegraphics[width=13.5 cm]{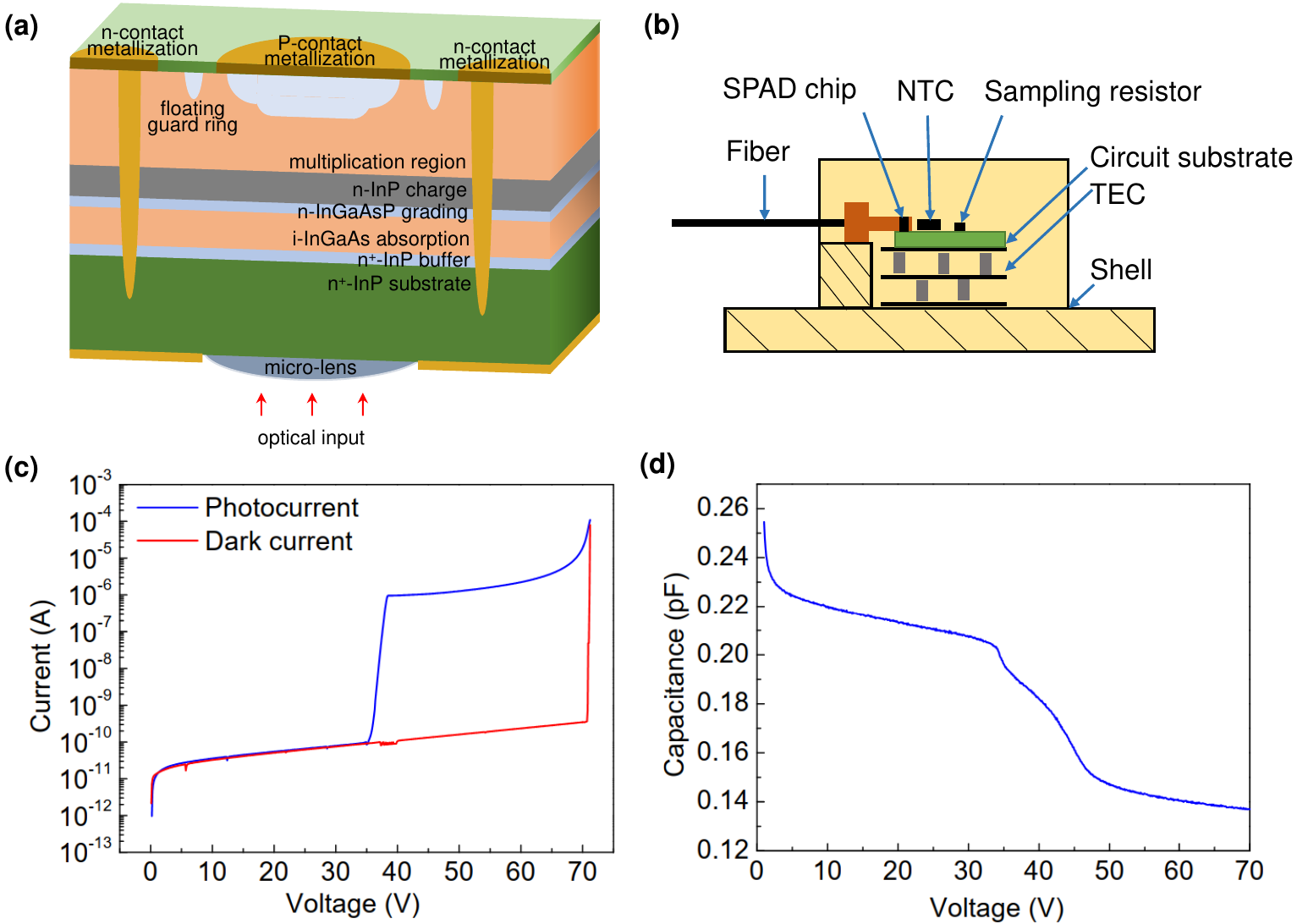}}
\caption{(a) SAGCM structure of the SPAD. (b) Structure of butterfly package. (c) Measured I-V curve of the SPAD. (d) Measured C-V curve of the SPAD.}
\label{figure2}
\end{figure*}

Follow the method\cite{ZJ16design}, we design and fabricate a SPAD chip using a separate absorption, grading, charge, and multiplication (SAGCM) structure as shown in Fig.~\ref{figure2} (a). Photons incident through the InP substrate and are absorbed in the InGaAs absorption layer. The InP charge layer is utilized to control the electric field distribution across the absorption and multiplication layers, enabling carriers in the absorption layer to reach the saturated drift velocity while suppressing tunneling currents. Simultaneously, a high avalanche probability is realized in the multiplication layer. The InGaAsP grading layer facilitates faster carrier transit across the heterojunction interface, thereby reducing time jitter. Additionally, a floating guard ring is designed to prevent edge pre-breakdown. The diameter of the active region is approximately 25 $\mu$m.

Compared to the previously proposed SPAD\cite{ZJ18MSWG}, we have made two major improvements. First, according to Telesca's theory\cite{Milan23CP}, thermal carriers from outside the active region could drift into it and contribute to dark counts. Therefore, an N-electrode with a groove structure is fabricated using wet-etching technology to isolate peripheral carriers, which leads to an optimized DCR performance. Second, a microlens with an 85 $\mu$m diameter is formed on the incident window using melted photoresist. This microlens focuses light from a 62.5 µm multimode fiber onto the active region, significantly enhancing the collection efficiency in Lidar applications.

The proposed SPAD chip is encapsulated in a butterfly package along with a mini-thermoelectric cooler (TEC). As shown in Fig.~\ref{figure2} (b), the TEC is soldered to a Corva alloy shell, onto which a ceramic substrate is glued. The SPAD chip is mounted on the substrate, and its electrodes are connected to the substrate via gold wire bonding. An avalanche sampling resistor and a negative temperature coefficient thermistor (NTC) are also welded on the substrate. The incident photons are aligned with the SPAD's sensitive area through a 62.5 µm metallized multimode fiber, which is fixed to the shell. The entire chamber is hermetically sealed and filled with nitrogen gas. The butterfly package SPAD has dimensions of 26 mm $\times$ 13 mm $\times$ 11 mm.

We characterized the SPAD at room temperature (300 K). Fig.~\ref{figure2} (c) shows the current-voltage (I-V) curve. The dark current near the breakdown voltage is as low as 0.4 nA, with the punch-through voltage and the breakdown voltage are 35 and 70 V, respectively. Fig.~\ref{figure2} (d) illustrates the capacitance-voltage (C-V) curve, showing a decreasing trend in total capacitance as the reverse bias voltage increases. Given a bias voltage of 70 V , the capacitance is approximately 0.137 pF.

\section{Monolithically Integrated Readout Circuit}

\begin{figure}[htp]
\centerline{\includegraphics[width=8 cm]{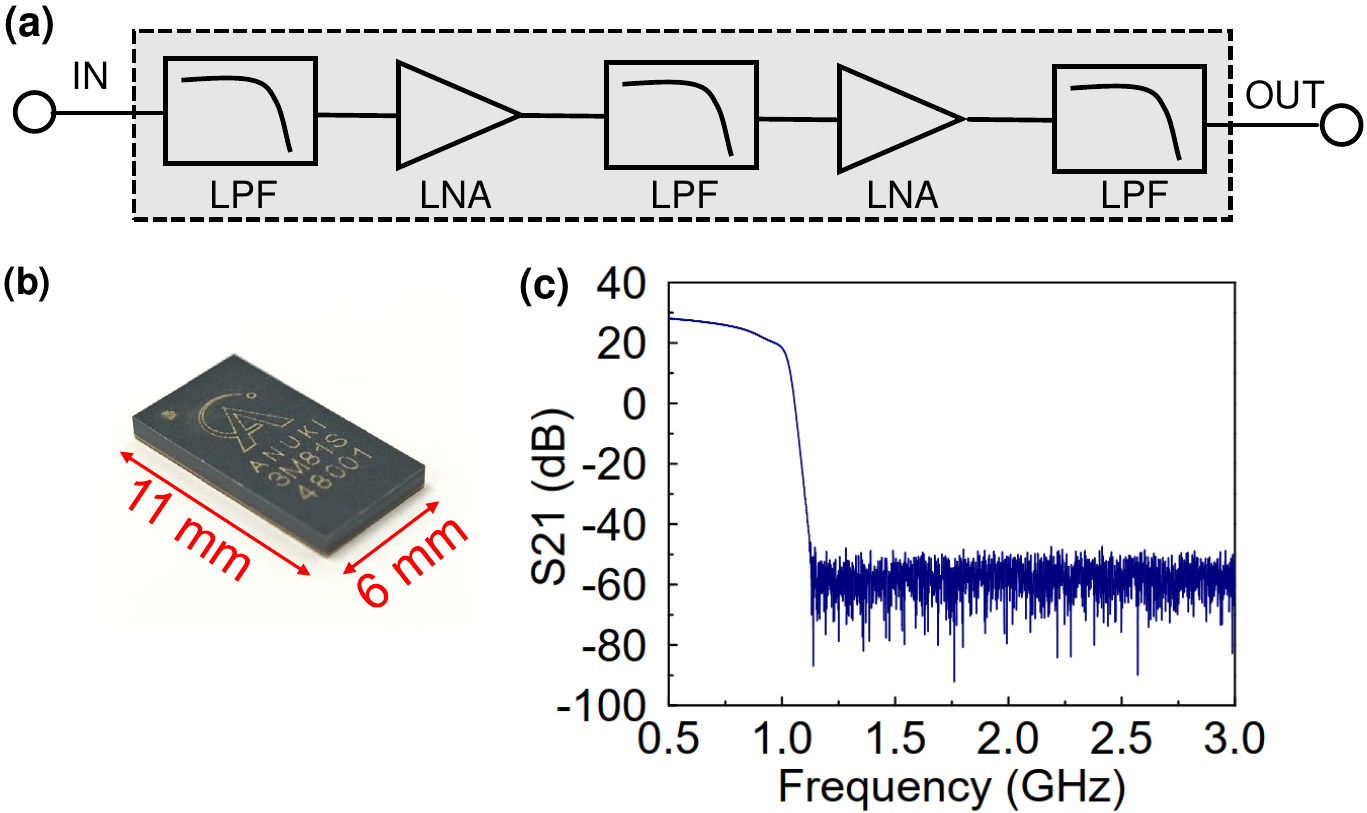}}
\caption{(a) Functional diagram of MIRC. LPF: low-pass filter. LNA: low-noise amplifier. (b) Photo of MIRC. (c) Measured S21 parameter of MIRC.}
\label{figure3}
\end{figure}

The capacitive response of the SPAD leads to an avalanche signal of only a few millivolts being superimposed onto the high-amplitude capacitive response signal of the sine wave gate, which can reach several volts. Consequently, the extraction of this weak avalanche signal is a significant challenge for high-frequency SWG SPDs. Several methods have been presented to effectively filter out the capacitive response signal while minimizing the impact on the avalanche signal, including microwave band-stop filters\cite{ZJ121G25}, ultra-narrowband interference circuits\cite{YZL23}, and the dual-anode SPAD technique\cite{Park19DA}. In contrast, low-pass filtering is a straightforward approach, which simutaneously removes high-frequency components of the avalanche signal, leading to increased time jitter. Nevertheless, it is particularly suitable for applications requiring a high degree of integration owing to its simplicity and ease of implementation.

\begin{figure*}[hbp]
\centerline{\includegraphics[width=17 cm]{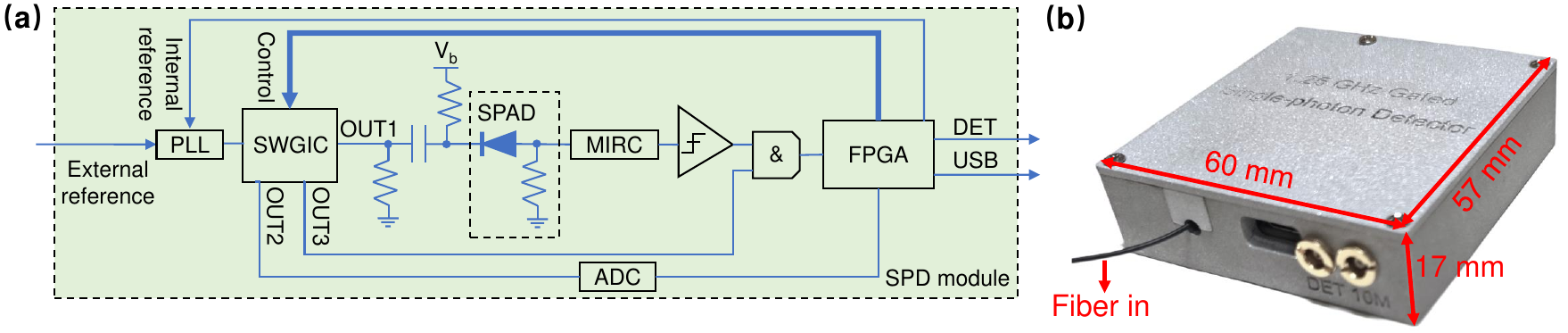}}
\caption{(a) Design diagram of the SPD module. PLL: phase-locked loop. SWGIC: sine wave gating integrated circuit. SPAD: single-photon avalanche diode. MIRC: monolithically integrated readout circuit. ADC: analog-to-digital converter. FPGA: field-programmable gate array. DET: detected signal. SPD module: single-photon detector module. (b) SPD module appearance.}
\label{figure4}
\end{figure*}

In our previous work, we developed a low-pass filtering MIRC chip with dimensions of 1.5 cm $\times$ 1.5 cm using low-temperature co-fired ceramics technology\cite{ZJ17LTCC}. In this paper, we present an enhanced version of the MIRC chip with a reduced size of 11 mm $\times$ 6 mm. The schematic diagram of the MIRC chip is depicted in Fig.~\ref{figure3} (a). The chip comprises three low-pass filters (LPFs) and two LNAs. The LPFs are fabricated using the super-high-performance integrated passive device technology\cite{ZCJ19IPD}, while the employed LNAs are commercial dies, each providing an approximate gain of 15 dB. All components are bonded onto a substrate and encapsulated using plastic sealing technology. A photograph of the MIRC chip is presented in Fig.~\ref{figure3} (b). 

The characterization results of the MIRC chip are shown in Fig.~\ref{figure3} (c). At frequencies below 1 GHz, the chip exhibits a gain of over 20 dB, with a maximum gain of 28 dB. Above 1 GHz, the gain sharply decreases, transitioning into signal suppression, which reaches 60 dB at 1.25 GHz. For higher frequencies, the chip maintain an average rejection ratio of $\sim$ 60 dB. This signifies that the proposed MIRC chip can effectively amplify the avalanche signal while filtering out the capacitive response signal from the sine wave gate. 

\section{SPD module}

Based on the SWGIC, SPAD, and MIRC, we implement a compact and fully functional 1.25 GHz AWG SPD module. The design schematic of the module is illustrated in Fig.~\ref{figure4} (a). A phase-locked loop (PLL), which can reference either an external or internal 10 MHz clock, generates the 1.25 GHz sine wave signal required for the SWGIC. The OUT1 of the SWGIC provides the amplified sine wave gate, which is AC coupled to the SPAD cathode. Furthermore, the avalanche current flows through a sampling resistor, producing an avalanche signal at the SPAD anode, which is superimposed on the sine wave gate capacitive response signal. This combined signal is then fed to the MIRC. The MIRC effectively extracts and amplifies the weak avalanche signals for further discrimination.

Due to the application of a low-pass filtering scheme, the time jitter of the avalanche signal increases slightly, resulting in a small fraction of avalanche events falling in the gate-off period. To mitigate the noise signal, a logic AND operation is performed using the coincidence signal from OUT3 of the SWGIC and the avalanche signal. By adjusting the relative delay between the coincidence signal and the sine wave gate, noise signals during the gate-off period are filtered out.

In the SPD module, a field-programmable gate array (FPGA) provides internal clock reference, controls SWGIC parameters, outputs a detection signal (DET), and communicates with a personal computer via a universal serial bus (USB). The voltage at the OUT2 of the SWGIC, which monitors the power of the sine wave gate, is digitized and sent to the FPGA, enabling real-time adjustments for maintaining the gate power at the desired setpoint. All components are integrated into a compact SPD module with dimensions of 6 cm $\times$ 5.7 cm $\times$ 1.7 cm. A photograph of the module is shown in Fig.~\ref{figure4} (b).

\begin{figure}[htbp]
\centerline{\includegraphics[width=9 cm]{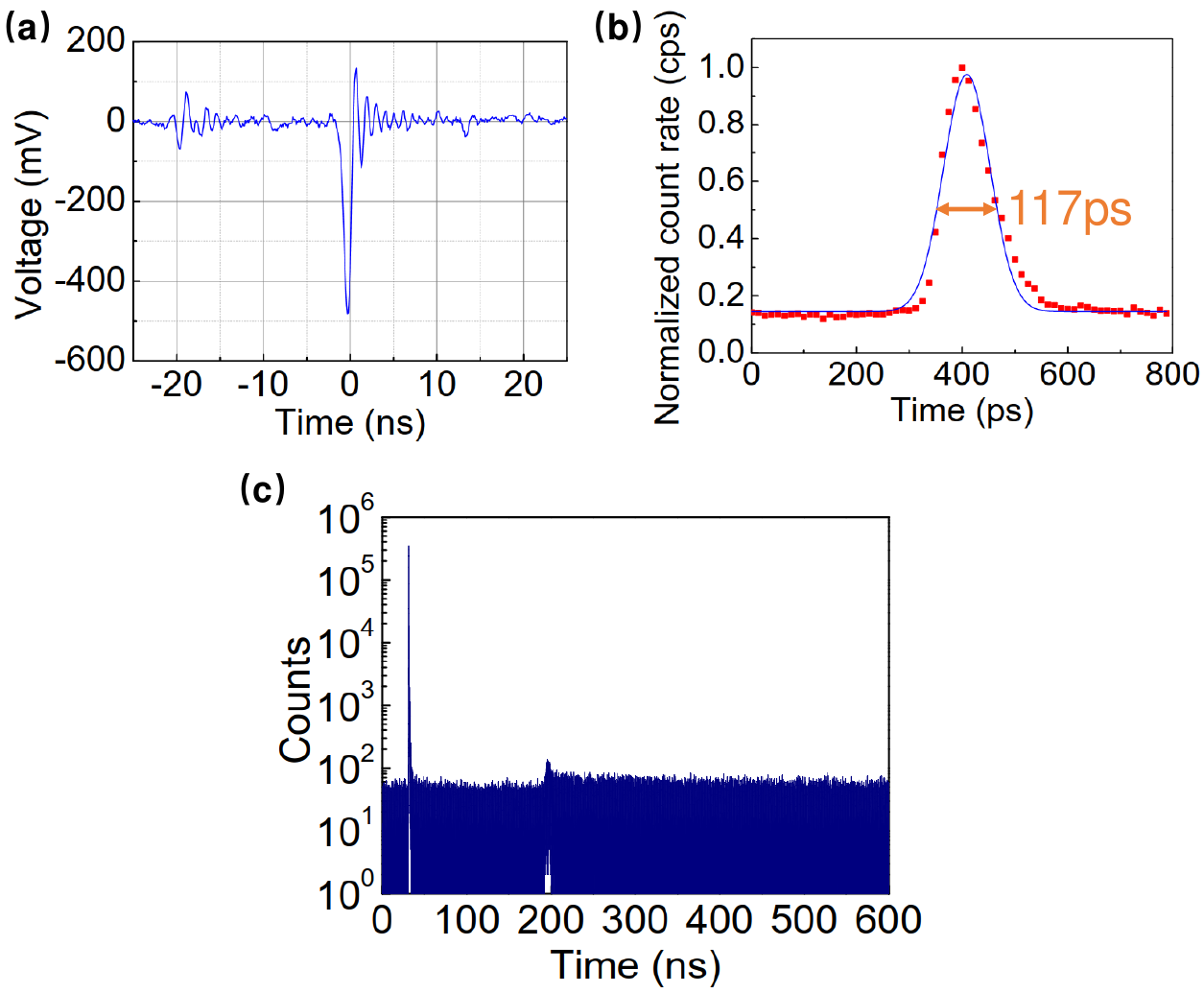}}
\caption{(a) A typical avalanche signal from the output of MIRC, acquired by an oscilloscope with 8 GHz bandwidth. (b) Effective gating width measurement at 40\% PDE and 238 K. (c) Typical detection histogram measured by TDC under 40\% PDE, 238 K temperature, 160 ns hold-off time and accumulation time of 1 minute. }
\label{figure5}
\end{figure}

The avalanche signal processed by the MIRC exhibits a high signal-to-noise ratio. When the sine wave gate amplitude is set to 20 V Vpp, the noise level remains below 100 mV, and the amplitudes of the avalanche signals exceed 200 mV. Fig.~\ref{figure5} (a) presents a typical avalanche signal with an amplitude of approximately 500 mV and a rise time of 2 ns. By scanning the delay between the laser pulses and the gating signal via the SWGIC, the effective gate width is measured to be 117 ps, as shown in Fig.~\ref{figure5} (b).

Subsequently, we characterize the SPD module following the standard method\cite{ZJ18MSWG}. A laser diode emits a pulse laser at a repetition frequency of 625 kHz with a pulse width of 70 ps. The laser is split into two channels: one for power monitoring and the other is further attenuated to 0.1 photons per pulse, which is connected to the SPD module. Detection events are recorded and analyzed using a time-to-digital converter (TDC). Fig.~\ref{figure5} (c) presents the typical TDC measurement results at the conditions of 40\% PDE and 1 minute accumulation time. In the figure, the first peak at 30 ns corresponds to photon detection events, with a FWHM timing jitter of 119 ps. The second peak at 190 ns represents afterpulse events, and the background noises represent dark counts.

\begin{figure}[htbp]
\centerline{\includegraphics[width=7 cm]{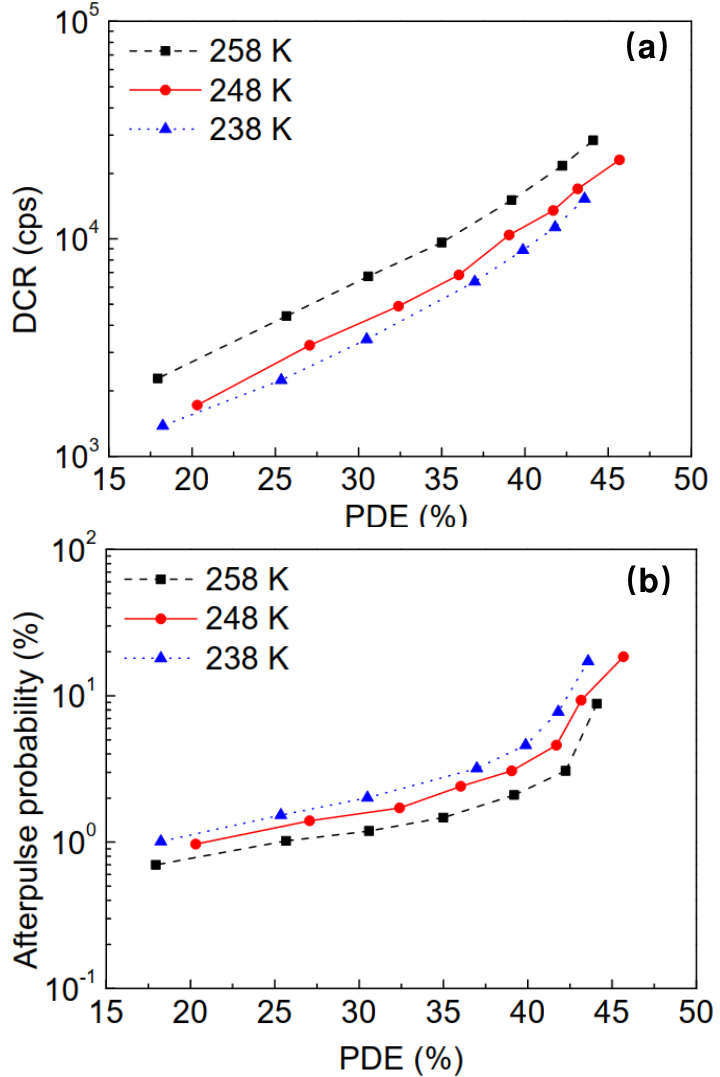}}
\caption{(a) DCR versus PDE and (b) afterpulse probability versus PDE at 20 V Vpp gate amplitude and 160 ns hold-off time.}
\label{figure6}
\end{figure} 

We characterize the performance of the SPD module at an operating point of 20 V Vpp gate amplitude and 160 ns hold-off time, which is suitable for QKD applications. The results are presented in Fig.~\ref{figure6}, with increasing temperature, the DCR rises, while the afterpulse probability decreases. At 248 K, the maximum PDE reaches 45.7\%, with a DCR of 23 kcps and an afterpulse probability of 18.5\%. For practical use, given a PDE of 40\%, a DCR of 9 kcps and an afterpulse probability of 4.6\% can be achieved at 238 K. Such performance satisfies the requirements of various integrated QKD systems. For other applications, parameters such as gate amplitude, hold-off time, and operating temperature can be flexibly optimized to achieve desired performance, depending on the specific requirements for the DCR, MCR, or other factors. 

Table~\ref{t1} summarizes the relevant works that focused on integrated high-frequency SWG SPDs. Notably, the proposed SPD module features the smallest size while maintaining full functionality. Moreover, the highest PDE is achieved at an acceptable noise level.

\begin{table*}[htp]
\centering
\caption{Performance comparison of integrated high-frequency SWG InGaAs/InP SPD in recent years.}
\label{t1}
\renewcommand\arraystretch{2}
%\begin{tabular}{lcccccccccc}
\begin{tabular}{|m{1.5cm}<{\centering}|m{2.5cm}<{\centering}|m{1cm}<{\centering}|m{2cm}<{\centering}|m{2cm}<{\centering}|m{1cm}<{\centering}|m{1cm}<{\centering}|m{2cm}<{\centering}|m{1cm}<{\centering}|m{1cm}<{\centering}|}
\hline
\hline
Paper   & Dimension ($cm^{3}$)    & Temp (K)     & PDE (\%)@1550 nm        & DCR ($gate^{-1}$)        & Dead time (ns)      & Pap (\%)    	        & Year    	\\
\hline
Ref. ~\cite{ZJ17LTCC}	 	& 12 $\times$ 7 $\times$ 5  		 & 223   	& 27.5        & 9.6 $\times$ $10^{-7}$    & 100       			& 9.1          			 & 2017 \\
%1.25 GHz sine wave gating InGaAs/InP single-photon detector with a monolithically integrated readout circuit
%\hline
Ref. ~\cite{ZJ18MSWG} 	 	& 13 $\times$ 8 $\times$ 4 		 	& 243   	& 30        		 & 3.5 $\times$ $10^{-7}$    & 100      				& 8.8          			 & 2018 \\
%Miniaturized high-frequency sine wave gating InGaAs/InP single-photon detector
%\hline
Ref. ~\cite{YZL23Compact}	 & 8.8 $\times$ 6 $\times$ 2 		 &  258   &  30       			 & 8.0 $\times$ $10^{-7}$   & 3       				& 2.4          			 & 2023 \\
%Compact InGaAs/InP Single-Photon Detector Module with Ultra-Narrowband Interference Circuits
%\hline
This work  	&  6 $\times$ 5.7 $\times$ 1.7  	 &  238   & 40        			& 7.2  $\times$ $10^{-6}$   & 160       				& 4.6\%          			 & 2024 \\
%\hline

\hline
\hline
\end{tabular}
\end{table*}

\section{Conclusion}

In conclusion, we have significantly reduced the size of the three critical components in high-frequency SWG SPDs by developing an SWGIC, a multimode fiber-coupled SPAD with a butterfly package, and an MIRC chip. Consequently, we have implemented the most compact high-frequency SWG SPD module to date, with dimensions of 6 cm $\times$ 5.7 cm $\times$ 1.7 cm. The SPD module integrates comprehensive functionalities, including gate amplitude modulation, gate amplitude monitoring, coincidence gate generation, phase adjustment, digital communication, and parameter configuration. The SPD module exhibits excellent performance of 40\% PDE, 9 kcps DCR, and 4.6\% afterpulse probability at an operation temperature of 238 K and a hold-off time of 160 ns, providing a practical and highly integrated solution for applications such as QKD and Lidar.

%\end{multicols}
\section*{Acknowledgment}
We would like to thank Anhui YUNTA Electronic Technologies Co., Ltd. for technical support. This work is supported by the Innovation Program for Quantum Science and Technology (2021ZD0300803, 2021ZD0300804) and the National Natural Science Foundation of China (62175227, 62405305).

\ifCLASSOPTIONcaptionsoff
  \newpage
\fi

%\section*{References}

\bibliography{SPD}
\bibliographystyle{IEEEtran}

\end{document}